\documentclass{aa}  

\usepackage{graphicx}
\usepackage{txfonts}
\usepackage{lipsum}
\usepackage{subcaption}         
\usepackage{lscape}            
\usepackage{placeins}          
\usepackage{hyperref}

\begin{document}

   \title{Evidence for the dynamical dark energy with evolving Hubble constant}

   \author{Yi-Ying~Wang\inst{1}
        \and Yin-Jie~Li\inst{1}
        \and Yi-Zhong~Fan\inst{1,2}\fnmsep\thanks{Corresponding author: yzfan@pmo.ac.cn}}

   \institute{Key Laboratory of Dark Matter and Space Astronomy, Purple Mountain Observatory, Chinese Academy of Sciences, Nanjing 210033, People's Republic of China
   \and School of Astronomy and Space Science, University of Science and Technology of China, Hefei, Anhui 230026, People's Republic of China}

   \date{Received December 09, 2025}

  \abstract 
   {Hubble constant tension, together with the recent indications of dynamical dark energy proposed from the Dark Energy Spectroscopic Instrument (DESI) baryon acoustic oscillation (BAO) measurements, poses significant challenges to the standard cosmological model. }
   {We investigate the possible redshift evolution of dark energy and the Hubble constant through a data-driven approach, and assess whether such evolution can alleviate the Hubble constant tension. } 
   {We perform a model-independent reconstruction of the dark-energy equation of state $w(z)$, jointly with an evolving Hubble constant $H_0(z)$. The analysis combines the DESI DR2 BAO dataset with multiple Type Ia supernova samples and evaluates the statistical preference for the reconstructed model using Bayesian evidence.}
   {The reconstructed $w(z)$ varies with redshift and exhibits two potential phantom crossings at $z\sim0.5$ and $z\sim1.5$. Meanwhile, $H_0$ decreases continually from local to high redshift, alleviating the Hubble constant tension effectively. The joint $w(z)$-$H_0(z)$ model is favored over the $w$CDM ($\Lambda$CDM) framework, with a logarithmic Bayes factor $\ln \boldsymbol{\mathcal B}= 5.04~(8.53)$. The results remain stable under different prior choices and dataset combinations. }
   {Our data-driven reconstructions suggest redshift evolution in both $w(z)$ and $H_0(z)$, offering a potential route to mitigate the Hubble constant tension. Future BAO measurements from Euclid and next-generation CMB experiments will provide critical tests of these results and bring deeper insights into the nature of dark energy and the evolution of cosmic expansion.}

   \keywords{cosmological parameters -- dark energy}

   \maketitle

\section{Introduction}
The standard cosmological paradigm, the cosmological constant $\Lambda$ and the Cold Dark Matter ($\Lambda$CDM) model, successfully explains most cosmological observations. However, the Hubble constant tension and possible evidence for dynamical dark energy pose significant challenges to this standard cosmological framework \citep{2022NewAR..9501659P, 2022JHEAp..34...49A, 2024hct..book.....D, 2025PDU....4901965D, 2026JCAP...01..009W} . Combining Cepheid-based distance measurements from the James Webb Space Telescope (JWST) and the Hubble Space Telescope (HST), \citet{2025ApJ...992L..34R} reported a local Hubble constant of $H_0 = 73.8 \pm 0.88 \, \rm km \, s^{-1} \, Mpc^{-1}$ via the distance-ladder method \citep{2021ApJ...908L...6R,2022ApJ...934L...7R}, which deviates from the $\Lambda$CDM prediction inferred from the cosmic microwave background (CMB) \citep{2020A&A...641A...6P} at the $\sim 6\sigma$ confidence level. Meanwhile, the recent Baryon acoustic oscillation (BAO) measurements from the Dark Energy Spectroscopic Instrument (DESI) Data Release 2 (DR2) \citep{2025PhRvD.112h3515A} prefer a dynamical dark energy model with CPL parameterization over $\Lambda$CDM at a $2.8-4.2\sigma$ confidence level when combined with CMB and type Ia supernovae (SNe Ia) data. A $\sim 5 \sigma$ level tension has been reported by \citet{2025PhRvD.112d3513S} when adopting different parameterization for $w(z)$.

Both the Hubble constant and dark energy govern the accelerating expansion of the Universe, describing its local expansion rate and the physical mechanism behind cosmic acceleration. The dark-energy equation of state (EoS), $w \equiv P/c^2\rho$, serves as a crucial diagnostic: $w =-1$ corresponds to vacuum energy within $\Lambda$CDM model. For dynamical dark energy, $w$ evolves with time and can be categorized into various models according to their trajectories in the $w$ phase space, including quintessence \citep{1988PhRvD..37.3406R}, phantom \citep{2002PhLB..545...23C}, quintom \citep{2005PhLB..607...35F}, $k$-essence \citep{PhysRevLett.85.4438} and others. For the phenomenological quintom model \citep{2006PhLB..634..101F}, an oscillating evolution of $w(z)$ naturally arises, accompanied with a varying $H_0$. Several studies suggest that late-time transitions in dark energy could alleviate the Hubble tension by reducing local $H_0$ values \citep{2020PhRvD.101j3517B, 2021Entrp..23..404D, 2025SCPMA..6880410P}. However, combining the latest DESI DR2 BAO and CMB data, a lower Hubble constant $H_0 = 63.7^{+1.7}_{-2.2} \, \rm km \, s^{-1} \, Mpc^{-1}$ was obtained \citep{2025PhRvD.112h3515A}, thereby exacerbating the Hubble tension. Furthermore, whether current data provide robust evidence for dynamical dark energy remains an open question \citep{2025PhRvD.112b3508G,2025MNRAS.542L..24O}.

To reconcile the discrepancies in Hubble constant measurements between early- and late-time cosmological observations, several studies have constrained $H_0$ by dividing the data into redshift bins. \citet{2020MNRAS.498.1420W} first reported an apparent trend of decreasing $H_0$ inferred from the individual lensing systems. A possible dynamical evolution of $H_0$ has since been investigated using a phenomenological function \citep{2021ApJ...912..150D, 2022Galax..10...24D, 2025JHEAp..4800405D}, modified definitions of the Hubble constant \citep{PhysRevD.103.103509, 2025ApJ...979L..34J}, and direct binned measurements \citep{PhysRevD.102.103525,2024EPJC...84..317M, 2024PDU....4401464O, 2025JCAP...03..026L}. Similarly, the EoS of dark energy $w(z)$ has been widely reconstructed in redshift bins without imposing an explicit functional form \citep{2012PhRvL.109q1301Z, 2017NatAs...1..627Z,  2025MNRAS.540.2253A}. Evidences for the oscillations in $w(z)$ have been found using non-parametric and analytical approaches \citep{2025NatAs.tmp..195G,2025arXiv250716970G}, respectively. Given the unknown essence of dark energy remains, such non-parametric, data-driven approaches provide a robust way to trace the evolution of $w(z)$.

In this work, we simultaneously investigate a model-independent dynamical dark energy and an evolving Hubble constant for the first time. We do not assume any specific forms for the dark energy and $H_0$, but instead reconstruct both functions directly from the data using a data-driven approach. The datasets we use include the DESI DR2 BAO measurements \citep{2025PhRvD.112h3515A}, SNe Ia observations from the PantheonPlus compilation \citep{2022ApJ...938..110B, 2022ApJ...938..113S}, the five-year Dark Energy Survey sample (DESY5) \citep{2024ApJ...973L..14D}, and the Union3 compilation \citep{2025ApJ...986..231R}. Without assuming any specific functional form, we employ Gaussian process to reconstruct $w(z)$ and $H_0(z)$ over the redshift range of $0<z<2.5$. The values of $w$ and $H_0$ are sampled in each redshift bin, enabling a complementary analysis that balances model flexibility and constraining power.

\section{Methods} \label{sec:methods}

Considering a flat universe ($\Omega_{\rm K} = 0$), the expansion rate is
\begin{equation}\label{eq:1}
	\frac{H(z)}{H_0(z)} = \bigr[\Omega_m (1+z)^3  + {\rm \Omega_{DE}} \frac{\rho_{\rm DE}(z)}{\rho_{\rm DE,0}}\bigr]^{1/2} , 
\end{equation}
where $\Omega_m$, $\rm \Omega_{DE}$ are the present-time matter and dark energy density parameters, respectively, and $H_0(z)$ represents the Hubble parameter evaluated in the corresponding-redshift bin. Here, $H_0(z)$ can be regarded as an effective, redshift-dependent normalization of the expansion rate, rather than a fundamental Hubble constant. As shown in \citet{PhysRevD.103.103509}, the Hubble constant can be represented as an integration constant from the Friedmann equation within the FLRW cosmology. Therefore, \autoref{eq:1} reduces to the FLRW framework only when $H_0(z)$ is a constant. The energy density $\rho_{\rm DE}$ that normalized to its present value evolves as 
\begin{equation}\label{eq:2}
	\frac{\rho_{\rm DE}(z)}{\rho_{\rm DE,0}} = {\rm exp}\bigg[{3 \int^z_0[1+w(z)]\frac{{\rm d}z'}{1+z'}}\bigg].
\end{equation}
The contributions from radiation and massive neutrinos are neglected for simplification. The BAO distance observables include the comoving distance $D_{\rm M}(z)$, the Hubble distance $D_{\rm H}(z)$, and the angle-average distance $D_{\rm V}(z)$, which can be written as
\begin{equation}\label{eq:3}
	\begin{aligned}
		D_{\rm M}(z) & = \int^{z}_{0} \frac{c \, {\rm d}z'}{H(z')}, \\
		D_{\rm H}(z) & = \frac{c}{H(z)},\\
		{\rm and~} D_{\rm V}(z) & = \big(z D_{\rm M}(z)^2 D_{\rm H}(z) \big)^{1/3}.
	\end{aligned}
\end{equation}
The pre-recombination sound horizon at the drag epoch is defined as $r_{\rm d} = \int^{\infty}_{z_{\rm d}} {c_{\rm s}(z)}/{H(z)} {\rm d} z$, where $c_{\rm s}$ is the sound speed in the primordial plasma. Following \citet{2023JCAP...04..023B} and \citet{2025PhRvD.112h3515A}, it can be approximated as 
\begin{equation}\label{eq:4}
	r_{\rm d} = 147.05 \, {\rm Mpc} \times \bigg( \frac{\Omega_{\rm b} h^2}{0.02236}\bigg )^{-0.13} \bigg(\frac{\Omega_{\rm m}h^2}{0.1432} \bigg )^{-0.23} \bigg(\frac{N_{\rm eff}}{3.04} \bigg )^{-0.1},
\end{equation}
where $h \equiv H_{0,z=0}/(100 \, \rm km \, s^{-1} \, Mpc^{-1})$ and $N_{\rm eff}$ is the effective number of neutrino species. In this work, only the variations in $\Omega_{\rm b}$ and $\Omega_{\rm m}$ are considered. The DESI DR2 BAO data cover $0.1<z<4.2$ using multiple tracers: the bright galaxy sample ($0.1<z<0.4$), luminous red galaxies ($0.4<z<1.1$), emission line galaxies ($1.1<z<1.6$), quasars ($0.8<z<2.1$) and Ly$\alpha$ forest ($1.77<z<4.16$). For the SNe Ia samples, the PantheonPlus compilation\footnote{We use the PantheonPlus compilation accompanied by the SH0ES Cepheids, obtained from \url{https://github.com/PantheonPlusSH0ES/}} \citep{2022ApJ...938..110B, 2022ApJ...938..113S} contains 1701 light curves of 1550 distinct SNe Ia covering $0.001<z<2.26$, the DESY5 sample \citep{2024ApJ...973L..14D} includes 1635 SNe Ia within $0.10<z<1.13$, and the Union3 compilation\footnote{The likelihoods of the DESY5 and Union3 compilations are calculated by the cosmological inference software \tt{cobaya} (\url{https://github.com/CobayaSampler}).} \citep{2025ApJ...986..231R} provides 2087 SNe Ia covering $0.001<z<2.26$. We use the observations that satisfy $z>0.01$. The luminosity distance is given by $D_{\rm L}(z)=(1+z)D_{\rm M}(z)$. 

We model the evolution of $w(z)$ and $H_0(z)$ over the redshift range $0<z<2.5$. The upper bound of $z=2.5$ is a conservative choice that encompasses all the observational datasets used in this analysis. The target range is divided into 29 bins of uniform width in scale factor ($a=1/(1+z)$) space \citep{2025NatAs.tmp..195G}. To reconstruct smooth and continuous functions of $w(z)$ and $H_0$ without assuming any explicit functional form, we employ Gaussian Process (GP) as a non-parametric statistical method. GP has been extensively applied in data-driven astronomy \citep{2010PhRvL.105x1302H, 2012PhRvD..85l3530S,2021ApJ...917...33L}, demonstrating powerful strength to sample the space of continuous functions. For prior assumptions, we adopt three types of kernels/methods to construct the covariance matrices $\bf \Sigma$: the $\rm Mat \acute{e} rn$ kernel $K_{\rm Mat \acute{e} rn}(\delta a, \, \nu, \, l)$, Gaussian kernel $K_{\rm Gaussian}(\delta a, \, l)$ and Horndeski theory \citep{1974IJTP...10..363H}. The general form of the covariance matrix is $\Sigma_{ij} = \sigma^2 K(|a_i - a_j|, l)$, where $\sigma$ represents the amplitude of the function, $a_i$ denotes the median point of the $i$-th bin in scale factor space, and $l$ governs the correlation length scale. For the Horndeski-based covariance, we adopt the prior covariance from Horndeski-based simulations \citep{PhysRevD.96.083509}, expressed as
\begin{equation}
	\begin{aligned}
		C(a_i, \, a_j) &= \sqrt{C(a_i) C(a_j)}R(a_i, \, a_j),\\
		C(a_i) &=0.05 + 0.8a_i^2, \\
		R(a_i, \, a_j) &= \exp{\big[-(|\ln a_i - \ln a_j|/0.3)^{1.2} \big]}.
	\end{aligned}
\end{equation}
Following \citet{2025NatAs.tmp..195G}, the covariance matrix $\bf \Sigma$ is calculated as
\begin{equation}
	{\bf \Sigma} = \big[({\rm \bf I} - {\rm \bf S})^{T} {\rm \bf C}^{-1} ({\rm \bf I} - {\rm \bf S})\big]^{-1},
\end{equation}
where $\rm \bf I$ is the identity matrix and $\rm \bf S$ is the transformation matrix that averages over 5 neighboring $w$ bins. The Horndeski-based covariance matrix predicts a variable $\sigma(w)$ ranging from 1.6 to 6.2 across bins. Accordingly, we adopt a constant $\sigma(w)=3$ for both the $\rm Mat \acute{e} rn$ and Gaussian kernels, in order to ensure a consistent comparison across different kernel and model assumptions. For the $H_0(z)$ reconstruction, we set $\sigma(H_0)=15$ as a weakly informative prior in the Bayesian inference,  ensuring that the reconstructed evolution of $H_0(z)$ is driven primarily by the data rather than by prior assumptions. The mean values are assumed to be $\mu(H_0) = 70$ and $\mu(w)=-1$, respectively. In this work, it should be noticed that Gaussian processes are employed as functional priors on $w(z)$ and $H_0(z)$, rather than as a direct interpolation technique. The correlation length $l$ is not optimised by maximizing or marginalizing GP likelihoods. Instead, we scan a range of fixed values and asses their performance by comparing the Bayesian evidence of the full cosmological model. We apply a flat prior on $\Omega_{\rm m}$ and a Gaussian prior on $\Omega_{\rm b}$, with $\Omega_{\rm m} \in [0.1, \, 0.5$] and $\Omega_{\rm b}h^2 = 0.02196\pm0.00063$ \citep{2024JCAP...06..006S}. Additional constrains require $H_0(z)$ and $w(z)$ remain within $[50, \, 90]$ and $[-6, \, 4]$, respectively.

\begin{figure*}[ht!]
	\centering
	\includegraphics[width=\linewidth]{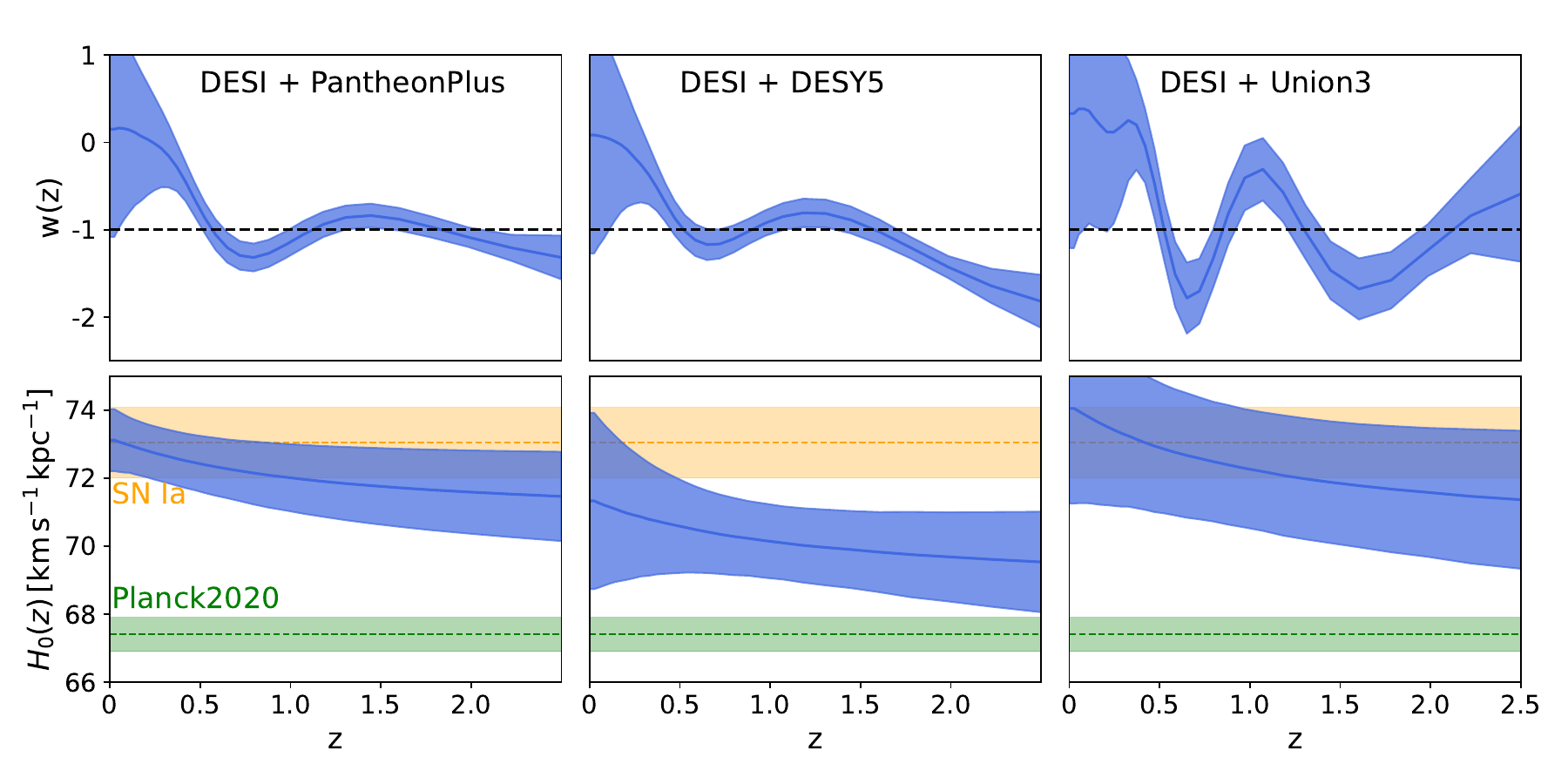}
	\caption{\small Reconstructed dark-energy EoS $w(z)$ and Hubble constant $H_0(z)$ from multiple datasets. The GP priors for $w(z)$ and $H_0(z)$ use $\rm Mat \acute{e} rn$-3/2 and Gaussian kernels, respectively. Panels in each column correspond to using the same datasets. Solid blue lines show the mean values of the $w(z)$ and $H_0(z)$ distributions. The shaded blue regions indicate the corresponding $68\%$ credible intervals. The dashed black lines marks $w=-1$. The orange and green bands denote  $H_0$ estimations from SNe Ia \citep{2022ApJ...934L...7R} and CMB \citep{2020A&A...641A...6P}, respectively.
	}
	\label{fig:1}
\end{figure*}

Since all of the components involved in the DESI BAO and SNe Ia measurements have been explicitly implemented, the likelihood functions in the Bayesian statistical framework can be written as
\begin{equation}
	\begin{aligned}
		\mathcal{L}_{\rm BAO}&=\prod^{N}_{i} {\rm exp} \left[ -\frac{1}{2} \left(\frac{f(x_i)-y_i}{\sigma_i} \right)^2 \right],\\
		\mathcal{L}_{\rm SNe} &= {\rm exp}\bigg[-\frac{1}{2}{\rm \Delta}{\boldsymbol{D}}^{T} C^{-1}_{\rm stat+syst} {\rm \Delta}{\boldsymbol{D}} \bigg],
	\end{aligned}
\end{equation}
where $(x_i, y_i)$ and $\sigma_i$ are the DESI BAO measurements and their uncertainties, $\boldsymbol{D}$ is the vector of SNe Ia distance modulus residuals, defined as ${\rm \Delta}D_i = \mu_i - \mu_{\rm model}(z_i)$, and $C_{\rm stat+syst}$ denotes the statistical and systematic covariance matrices. To balance the efficiency and accuracy, we employ the nested sampling algorithm implemented in {\tt pymultinest} \citep{2016ascl.soft06005B} and set 2000 live points during Bayesian analysis. 

\section{Results\label{sec:Results}}

\begin{figure*}[ht!]
	\centering
	\includegraphics[width=1.\linewidth]{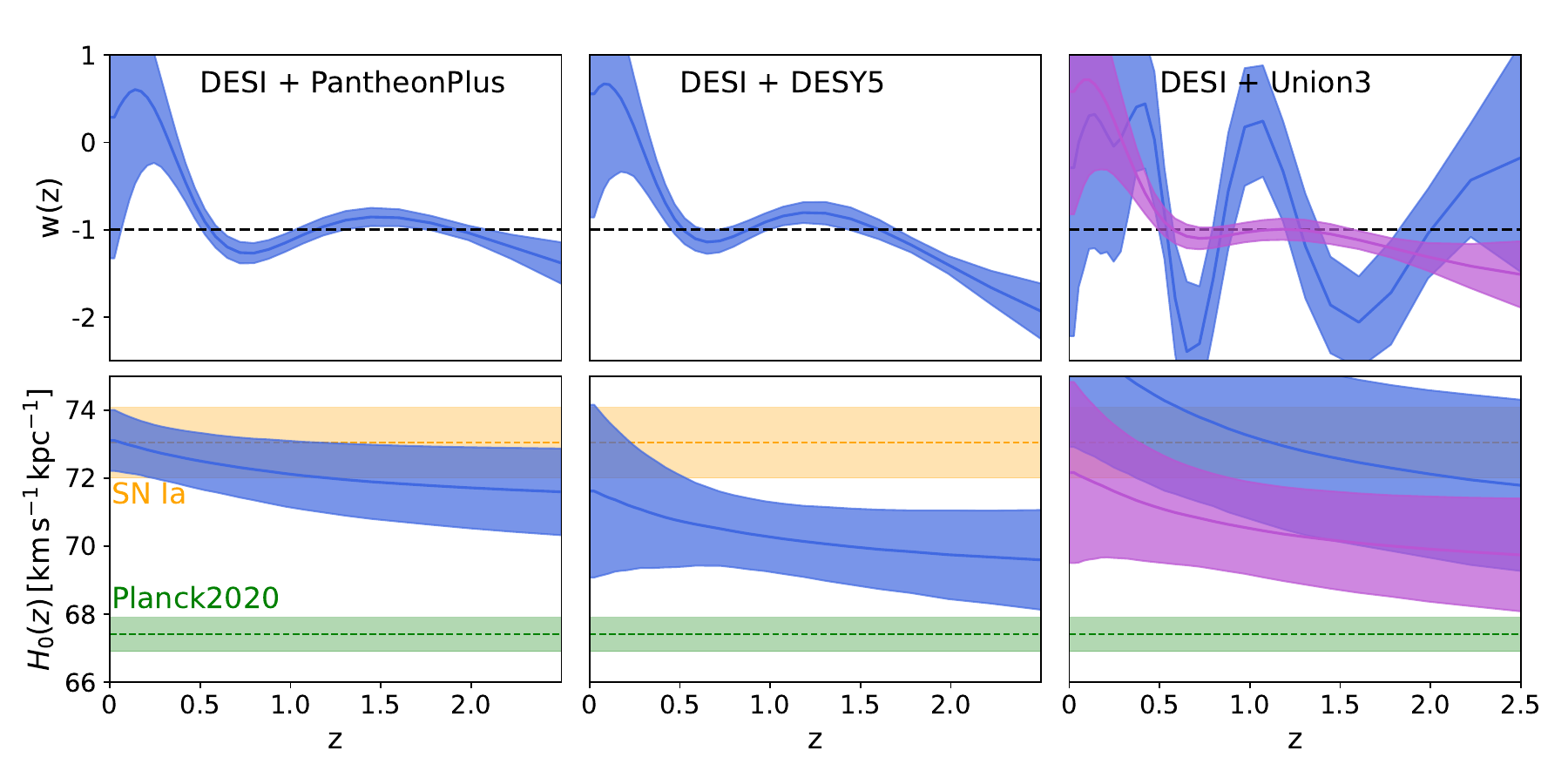}
	\caption{\small Similar to \autoref{fig:1}, but showing the posterior reconstruction of $w(z)$ obtained by adopting a Gaussian kernel to define the GP prior covariance. In the right panel, the blue region represents the results with a maximum Bayes factor with $l=0.05$. The purple region corresponds to $l=0.2$, assuming the same scale length as in the left and middle panels. A smaller $l$ ($l<0.01$) in the DESI + Union case may yield a higher Bayesian evidence, but the resulting $w(z)$ becomes too flexible to provide clear constraint. Therefore, the right panel only shows the results for $l=0.05$ and $l=0.2$.}
	\label{fig:2}
\end{figure*}

\begin{figure*}[ht!]
	\centering
	\includegraphics[width=1.\linewidth]{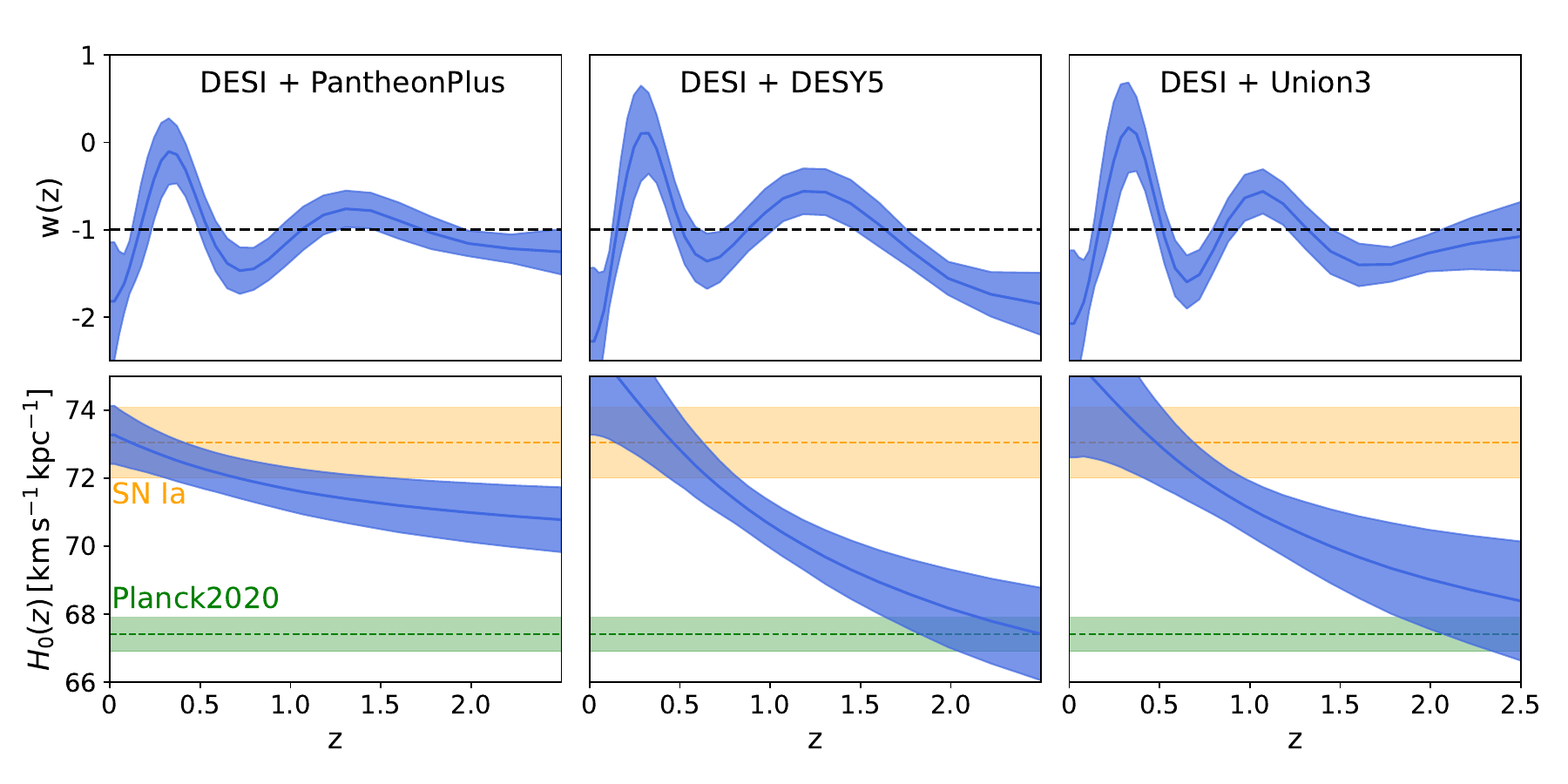}
	\caption{\small Similar with \autoref{fig:1}, but using the Horndeski theory to derive the $w(z)$ prior. The results in the left panel use a Gaussian kernel with $l=2.0$ to derive the $H_0(z)$ priors, while the others use $l=1.0$ to obtain the maximum bayes factors.}
	\label{fig:3}
\end{figure*}

The evolutions of $w(z)$ and $H_0(z)$ are reconstructed simultaneously. Each is divided into 29 bins. The priors of {\boldmath $w$} $=(w_1, \, ..., \, w_{29})$ and {\boldmath $H_0$} = $(H_{0,1}, \, ..., \, H_{0,29})$ are generated either from the Horndeski theory or from the GP kernel functions, expressed as $\pi(${\boldmath $w$}$|\mu, \sigma, l)$ and $\pi(${\boldmath $H_0$}$|\mu, \sigma, l)$, respectively. Besides $\sigma$, the length scale $l$ acts as  another hyper parameter that regulates the smoothness of reconstructed functions. Larger $l$ values enforce stronger correlations, stiffening the model that will ignore the details, while smaller $l$ values soften the model but may risk overfitting. We compare Bayesian evidences across several configurations with different $l$ values and find that $H_0(z)$ shows a clear decreasing trend. This preference for a smoother function for $H_0(z)$ leads us to adopt a Gaussian kernel with a relatively large length scale. Detailed comparisons of the scale length are provided in \autoref{sec:conclusion}. 

\autoref{fig:1} illustrates the reconstructed $w(z)$ (top panels) and $H_0(z)$ (bottom panels) derived from various data combinations. The priors of {\boldmath $w$} are generated using the $\rm Mat \acute{e} rn$-3/2 kernel with $l=2.5$ and $\nu=2/3$ for the DESI + PantheonPlus and DESI + DESY5 datasets, and $l=0.9$ for the DESI + Union3 case. The priors of {\boldmath $H_0$} are generated using the Gaussian kernel with $l=2.0$. These configurations yield the highest Bayesian evidence among all tested $l$ values, rather than being fitted through a Gaussian process or marginal likelihood optimization. In all cases, $w(z)$ exhibits clear redshift evolution with multiple peaks, while $H_0(z)$ demonstrates a significant decreasing trend from $\sim 73$ to $\sim 70 \, \rm km~s^{-1}~Mpc^{-1}$, approaching the CMB-inferred value. To quantify the model preference for dynamical dark energy and an evolving Hubble constant, we compute the Bayes factor, $\boldsymbol{\mathcal{B}} = \frac{ P(\boldsymbol{y}|\mathcal{M}_{w(z) {\rm CDM}+H_0(z)}) }{ P(\boldsymbol{y}|\mathcal{M}_{w{\rm CDM}}) }$. The priors of the $w$CDM model are consistent with those described previously. From left to right in \autoref{fig:1}, we obtain $\ln \boldsymbol{\mathcal{B}} = 4.18$, $3.39$, and $2.13$, respectively. Compared with $\Lambda$CDM models, Bayes factors give $\ln \boldsymbol{\mathcal{B}} = 0.69$, $7.67$, and $1.54$, respectively. As pointed out by \citet{1939thpr.book.....J}, $\ln \boldsymbol{\mathcal{B}} >2.3$ ($\ln \boldsymbol{\mathcal{B}} >3.5$) indicates a strong (very strong) preference for one model over another, and $\ln \boldsymbol{\mathcal{B}} >4.6$ is regarded as decisive evidence. Thus, the DESI + PantheonPlus and DESI + DESY5 datasets provide a very strong and strong support for the joint $w(z)$-$H_0(z)$ model. 

As a robustness test, the results using Gaussian kernels and Horndeski theory are shown in \autoref{fig:2} and \autoref{fig:3}, respectively. The priors of {\boldmath $H_0$} are generated using the Gaussian kernel with $l=2.0$, except for the middle and right panels in \autoref{fig:3}, which use a Gaussian kernel with $l=1.0$. The Gaussian kernel yields smoother reconstructions with larger length scale, whereas the $\rm Mat \acute{e} rn$ kernel allows greater flexibility. The Horndeski prior produces the most flexible $w(z)$ evolution, capturing more distinct features in the observations. In \autoref{fig:2}, the blue regions corresponds that the priors of {\boldmath $w$} are generated by Gaussian kernel with $l=0.2$ for DESI + PantheonPlus and DESI + DESY5, and $l=0.05$ for DESI + Union3 scenarios. The resulting Bayes factors relative to $w$CDM ($\Lambda$CDM) model are $\ln \boldsymbol{\mathcal{B}} = 3.74$ (0.23), $2.96$ (7.24), and $2.86$ (2.27), respectively, again showing very strong preference for DESI + PantheonPlus and strong preference for other datasets. The purple regions in the right panels represent reconstructions with $l=0.2$, yielding consistent shapes compared with other scenarios using different datasets but a smaller $\ln \boldsymbol{\mathcal{B}} = 0.64$. For the results using the Horndeski prior shown in \autoref{fig:3} , Bayes factors relative to $w$CDM ($\Lambda$CDM) model are $\ln \boldsymbol{\mathcal{B}} = 5.04$ (1.55), $4.25$ (8.53), and $3.06$ (2.47) for differ datasets, respectively. Notably, the DESI + PantheonPlus combination provides a decisive evidence ($\ln\boldsymbol{\mathcal{B}} > 4.6$) favoring the $w(z)$-$H_0(z)$ model.

All nine scenarios, constructed with different dataset combinations and prior assumptions, consistently indicate that both the dark-energy EoS $w(z)$ and the Hubble constant $H_0$ can evolve with redshift. Compared with \autoref{eq:1} and \autoref{eq:2}, the evolved $H_0$ suggests that dynamical dark energy alone may not be sufficient to fully explain the Hubble constant tension, potentially indicating the presence of new physics beyond the standard cosmological model. Several deviations from $\Lambda$CDM model appear at distinct epochs. The reconstructed $w(z)$ crosses into the phantom regime ($w<-1$) around $z\approx0.5$, consistent with previous non-parametric DESI BAO analyses \citep{w4c6-1r5j}. It returns to the quintessence regime around $z\approx1.0$, and then crosses again into the phantom region near $z\approx1.5$. Particularly, all reconstructions show a preference for $w(z)$ exhibiting two phantom crossings. Furthermore, no clear non-monotonic behavior is observed in $H_0(z)$. Instead, all results present a gradual decline from local to higher redshift, with different amplitudes depending on the dataset and prior choice. The DESI + PantheonPlus case with Gaussian kernel yields the smallest variation, from $73.11_{-0.85}^{+0.90}$ to $71.60_{-1.26}^{+1.27} \, \rm km \, s^{-1} \, Mpc^{-1}$, while the DESI + DESY5 case with Horndeski theory gives the largest one, from $75.81_{-2.53}^{+3.39}$ to $67.41_{-1.12}^{+1.36} \, \rm km \, s^{-1} \, Mpc^{-1}$. Compared with the results based on the classical CPL parameterization \citep{2025PhRvD.112h3515A}, a lower $H_0$ value is typically obtained when $w_0>-1$, as a consequence of parameter degeneracy inherent in the CPL model \citep{2022JCAP...04..004L}. Differently, our result alleviates the Hubble constant tension through the downward evolution of $H_0(z)$, enabled by a dynamical dark energy with greater freedom than the CPL parameterization.

\begin{figure*}[ht!]
	\centering
	\includegraphics[width=1.\linewidth]{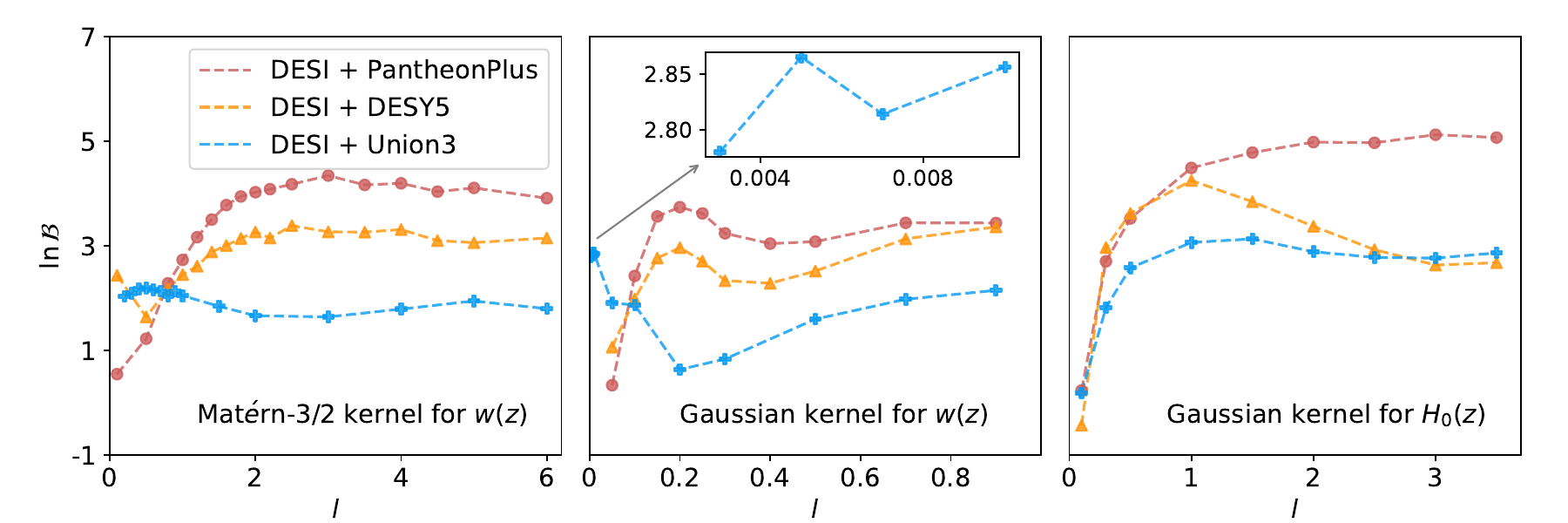}
	\caption{\small Comparisons with different scale length under various prior assumptions. The red, orange, and blue scatters represent the Bayes factors relative to $w$CDM models using DESI combined with PantheonPlus, DESY5, and Union3 datasets, respectively.}
	\label{fig:lnZ}
\end{figure*}

\section{Conclusions and discussions} \label{sec:conclusion}
In this work, we reconstruct the dynamical dark-energy EoS $w(z)$ together with an evolving Hubble constant $H_0(z)$ by non-parametric methods. Across multiple dataset combinations and prior assumptions, we find consistent evolution patterns: $w(z)$ evolves with redshift and emerges two potential phantom crossing, and $H_0(z)$ decreases smoothly toward high redshift. In particular, the DESI + DESY5 scenario with the Horndeski prior gives the largest downward, $\Delta H_0 \sim 8.4 \, \rm km \, s^{-1} \, kpc^{-1}$, reaching $H_0 \sim 67.4 \, \rm km \, s^{-1} \, Mpc$. Compared with $w$CDM model and $\Lambda$CDM model ($w = -1$), the Bayes factors are $\ln \boldsymbol{\mathcal{B}} = 4.25$ and $8.53$, respectively, providing a promising solution to relieve the Hubble constant tension. Furthermore, the maximum Bayes factor among all scenarios compared with $w$CDM model reaches $\ln \boldsymbol{\mathcal{B}} = 5.04$, establishing decisive evidence for $w(z)$-$H_0(z)$ model.

The above results and discussions depend on the choice of the characteristic scale length $l$. To clarify the selection of $l$ in each case, we compare the Bayes factors relative to the $w$CDM model for different scale factor $l$ across multiple scenarios. As shown in \autoref{fig:lnZ}, the panels from left to right display the variations of $\ln \boldsymbol{\mathcal{B}}$ in the reconstructions of $w(z)$ using $\rm Mat \acute{e} rn$-3/2 and Gaussian kernels, and of the $H_0(z)$ using Gaussian kernels. In the left and middle panels, the priors of $H_0(z)$ are generated with a Gaussian kernel of $l=2.0$, while in the right panel, the priors of $w(z)$ are derived from the Horndeski theory and the priors of $H_0(z)$ are constructed with a Gaussian kernel using variable $l$. The variations of $\ln \boldsymbol{\mathcal{B}}$ in the middle panel exhibit a moderate increase with larger $l$, which may result from the reduced effective degrees of freedom introduced by smoother Gaussian-kernel priors. The maximum values of $\ln \boldsymbol{\mathcal{B}}$ correspond to the results presented in \autoref{sec:Results}. Detailed parameter estimation results are listed in \autoref{Tab:app}.

\begin{figure}[ht!]
	\centering
	\includegraphics[width=0.9\linewidth]{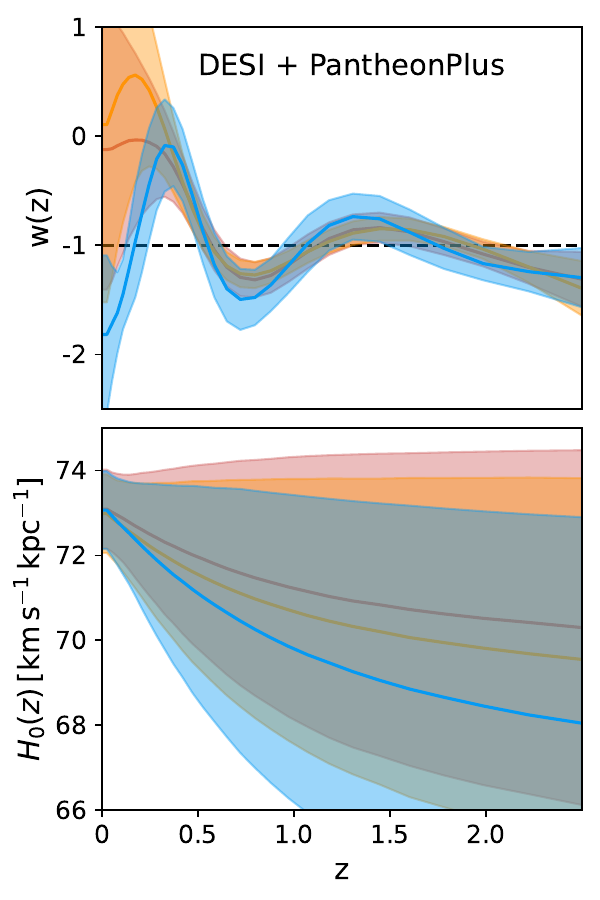}
	\caption{\small The reconstructions of $w(z)$ and $H_0(z)$ using the DESI + PantheonPlus datasets. The prior of $H_0(z)$ uses the Gaussian kernel with $l=1.5$. he shaded  regions indicate the corresponding $68\%$ credible intervals. The red, orange and blue regions correspond to $w(z)$ priors generated from the $\rm Mat \acute{e} rn$-3/2 kernel ($l=2.5$), the Gaussian kernel ($l=0.2$), and the Horndeski theory, respectively. }
	\label{fig:4}
\end{figure}

Given that the measurements of BAO relies on the pre-recombination sound-horizon scale and the degeneracy between $r_{\rm d}$ and $H_0$, we examine  the impact of $r_{\rm d}$ on the constructions of $w(z)$ and $H_0(z)$. Instead of adopting the fixed relation in \autoref{eq:4}, we assume $r_{\rm d}$ to be a free parameter with a uniform prior of $[120,\, 180] \, \rm Mpc$. \autoref{fig:4} shows the results using the DESI + PantheonPlus datasets. The reconstructed $w(z)$ remains consistent with previous cases, and $H_0(z)$ continues to exhibit the same decreasing trend. As expected, the uncertainties in  $H_0(z)$ increase because BAO measurements constrain the product $H_0 r_{\rm d}$ more tightly than the individual parameters \citep{2025PhRvD.112h3515A}. For example, under the Horndeski prior, $H_0$ declines from  $73.06^{+0.92}_{-0.91}$ to $68.04^{+4.86}_{-3.65} \, \rm km \, s^{-1} \, kpc^{-1}$. Three different prior assumptions are used, including the $\rm Mat \acute{e} rn$-3/2 kernel, the Gaussian kernel with $l=0.2$, and the Horndeski theory. The Bayes factors relative to the $w$CDM model are $\ln \boldsymbol{\mathcal{B}} = 3.17$, $2.66$, and $4.02$, respectively, corresponding to two strong and one very strong  preference for the $w(z)$-$H_0(z)$ framework.

\begin{figure}[ht!]
	\centering
	\includegraphics[width=0.9\linewidth]{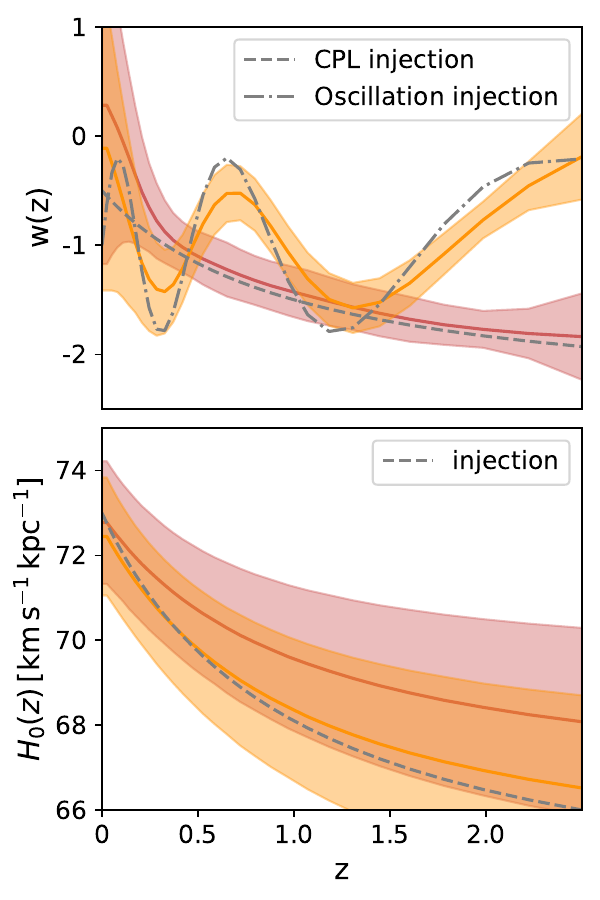}
	\caption{\small The reconstructions of $w(z)$ and $H_0(z)$ based on simulated injection. The grey lines denote the injected models. For $w(z)$, the dashed grey line represents the CPL parameterization with $w(z) =-0.5-2(1-1/(1+z))$ and the dash-dot grey line shows the oscillation model with $w(z) = 0.8\sin[20(1-1/(1+z))]-1$. For simplicity, the other injected parameters are fixed with $\Omega_m = 0.3$, $\omega_b h^2 = 0.02196$, and $M_b = -19.5$. For $H_0(z)$, the dashed grey line corresponds $H_0(z) = 73-9.8(1-1/(1+z))$. The shaded regions indicate the corresponding $90\%$ credible intervals. The prior of $H_0(z)$ uses the Gaussian kernel with $l=2.0$, and the prior of $w(z)$ is generated from the $\rm Mat \acute{e} rn$-3/2 kernel ($l=2.5$). }
	\label{fig:6}
\end{figure}

The reconstruction of $w(z)$ using both $\rm Mat \acute{e} rn$-3/2 and Gaussian kernels shows consistent results with the Horndeski prior, particularly the evidence of two phantom crossings. Based on independent analyses, \citet{2025NatAs.tmp..195G} and \citet{2025arXiv250716970G} have reported oscillatory features in $w(z)$ consistent with our results. To validate the reconstruction method, we perform the same procedure using the simulated data with observational uncertainties matching those of PantheonPlus and DESI DR2. As shown in \autoref{fig:6}, both the CPL parameterization and the oscillation model can be accurately reconstructed using the same kernel function. For the scenario using the CPL parameterization, no pseudo oscillation arise, demonstrating the robustness of our method and the reliability of the physical transition signatures identified in our results.

As one of the most general scalar-tensor theories involving second-order field equations, the Horndeski framework has been widely investigated for its potential to alleviate the Hubble tension by predicting a large $H_0$ \citep{2022PhRvD.106l4051P, 2023PhRvD.108b4012B, 2024EPJC...84..220T}. After defining the effective Hubble parameters through the expansion rate of $\Lambda$CDM and $w$CDM models, \citet{2025PDU....4801847M} and \citet{2025ApJ...994L..22J} proposed that a running $H_0$ naturally arises from evolving dark energy. Because phantom-crossing behavior can drive a faster expansion phase leading to large $H_0$, the concept of an evolved $H_0(z)$ is theoretically motivated. Furthermore, our reconstructions of $w(z)$ are compatible with various dynamical dark-energy models. If dark energy can possess multiple internal degrees of freedom, such as multi-field models \citep{2005PhRvD..71d7301H, 2005PhLB..608..177G, 2008JCAP...03..013C}, modified vacuum models \citep{2000PhRvD..62h3503P, 2006PhRvD..73b3513C}, and interacting dark-energy models \citep{2000PhRvD..61h3503B}, phantom crossing can be generated properly.

Beyond traditional approaches, many novel astrophysical probes covering a broad range of redshift have been proposed to measure the cosmological parameters and the dark-energy EoS. For instance, the spectral peak of gamma-ray burst (merger events via gravitational-wave observations) can serve as a tracer (standard sirens) of cosmic expansion \citep{2004ApJ...612L.101D,2017Natur.551...85A,2023ApJ...943...13W,2024ApJ...976..153L}, 
the dispersion measures of fast radio bursts provide independent constraints on $H_0$ and $w(z)$ \citep{2014PhRvD..89j7303Z, 2025ApJ...981....9W}, and the propagation speed of gravitational waves from multi-messenger events can test large classes of scalar-tensor and dark-energy theories \citep{2017PhRvL.119y1304E}. In addition to $w(z)$ and $H_0$, several other cosmological parameters have been modeled as redshift-dependent functions to explain current cosmological tensions \citep{2022PhRvD.106d1301O, 2026ApJ...996L...5F, 2024PDU....4401464O}. Looking forward, next-generation CMB experiments (CMB-S4, the Simons Observatory, and LiteBIRD) \citep{2022arXiv220308024A, 2019BAAS...51g.147L, 2023PTEP.2023d2F01L}, together with high-precision BAO surveys such as Euclid \citep{2025A&A...697A...1E}, will substantially tighten cosmological constraints and provide crucial insights into the physics of dark energy and cosmic expansion.

\begin{acknowledgements}
We thank the anonymous referee for helpful comments and suggestions. This work is supported in part by NSFC under grants of No. 12233011, the Postdoctoral Fellowship Program of CPSF (No. GZB20250738), and the Jiangsu Funding Program for Excellent Postdoctoral Talent (No. 2025ZB209).

\end{acknowledgements}

\bibliographystyle{aa.bst} 
\bibliography{Refs.bib} 

\begin{appendix}

\onecolumn

\renewcommand{\arraystretch}{1.4}
\begin{sidewaystable*}[htbp]
	
\section{The results of parameter estimations for various cosmological models}\label{sec:app}

This appendix presents the posterior distributions of the cosmological parameters and the corresponding Bayesian evidences for all scenarios discussed in the main text. For the reconstructions of $H_0(z)$, we show the values at $z\sim0.01$ and $z\sim0.36$. For $w(z)$, the specific and local extreme values are listed.
	\centering
	\begin{tabular}{l|c@{\hspace{0.1em}}c@{\hspace{0.1em}}c@{\hspace{0.1em}}c@{\hspace{0.1em}}c@{\hspace{0.1em}}c@{\hspace{0.1em}}c@{\hspace{0.1em}}c@{\hspace{0.5em}}c@{\hspace{0.5em}}c@{\hspace{0.5em}}c}
		\hline
		\multicolumn{12}{c}{Posterior distributions in 68\% credible intervals for multi cases}\\
		\hline
		Datasets & Cases & $H_0(z\sim0.01)$ & $H_0(z\sim2.35)$ & $w(z\sim0.01)$ & $w(z\sim0.62)$ & $w(z\sim1.02)$ & $w(z\sim1.52)$ & $w(z\sim2.35)$ & $\Omega_{\rm m}$ & $\Omega_{\rm b}h^2$ & $\ln(E)$ \\
		&  & \multicolumn{2}{c}{$(\rm km \, s^{-1} \, Mpc^{-1})$} &  &  &  &  &  &  &  &  \\
		\hline
		&M &$73.12^{+0.91}_{-0.85}$ &$71.46^{+1.31}_{-1.35}$ &$0.15^{+1.23}_{-1.16}$ &$-1.20^{+0.17}_{-0.17}$ &$-1.06^{+0.15}_{-0.15}$ &$-0.88^{+0.13}_{-0.13}$ &$-1.32^{+0.25}_{-0.25}$ &$0.35^{+0.02}_{-0.02}$ &$0.02193^{+0.00053}_{-0.00050}$ &-744.96 \\
		&G &$73.11^{+0.90}_{-0.85}$ &$71.60^{+1.27}_{-1.26}$ &$0.29^{+1.61}_{-1.45}$ &$-1.20^{+0.12}_{-0.12}$ &$-1.06^{+0.10}_{-0.10}$ &$-0.86^{+0.10}_{-0.10}$ &$-1.38^{+0.24}_{-0.24}$ &$0.35^{+0.02}_{-0.02}$ &$0.02191^{+0.00051}_{-0.00051}$ &-745.40 \\
		{D+P} &H &$73.27^{+0.85}_{-0.88}$ &$70.78^{+0.95}_{-0.98}$ &$-1.82^{+0.68}_{-0.69}$ &$-1.39^{+0.29}_{-0.28}$ &$-0.98^{+0.25}_{-0.24}$ &$-0.90^{+0.21}_{-0.20}$ &$-1.25^{+0.26}_{-0.26}$ &$0.34^{+0.02}_{-0.02}$ &$0.02193^{+0.00054}_{-0.00052}$ &-744.11 \\
		&$w$CDM &$72.18^{+0.77}_{-0.79}$ &$-$ &$-0.96^{+0.03}_{-0.03}$ &$-$ &$-$ &$-$ &$-$ &$0.30^{+0.01}_{-0.01}$ &$0.02246^{+0.00053}_{-0.00058}$ &-749.14 \\
		&$\Lambda$CDM & $72.77^{+0.54}_{-0.55}$ &$-$ &$-$ &$-$ &$-$ &$-$ &$-$ &$0.31^{+0.01}_{-0.01}$ &$0.02228^{+0.00055}_{-0.00056}$ &-745.65 \\
		\hline
		&M &$71.33^{+2.59}_{-2.61}$ &$69.53^{+1.48}_{-1.38}$ &$0.08^{+1.36}_{-1.11}$ &$-1.17^{+0.18}_{-0.17}$ &$-0.85^{+0.15}_{-0.15}$ &$-1.02^{+0.14}_{-0.14}$ &$-1.82^{+0.30}_{-0.32}$ &$0.37^{+0.04}_{-0.04}$ & $0.02198^{+0.00050}_{-0.00051}$ &-840.89 \\
		&G &$71.62^{+2.54}_{-2.58}$ &$69.59^{+1.47}_{-1.35}$ &$0.55^{+1.41}_{-1.44}$ &$-1.14^{+0.14}_{-0.12}$ &$-0.84^{+0.11}_{-0.11}$ &$-1.00^{+0.11}_{-0.11}$ &$-1.93^{+0.32}_{-0.33}$ &$0.37^{+0.04}_{-0.04}$ &$0.02196^{+0.00052}_{-0.00050}$ &-841.32 \\
		{D+D} &H &$75.81^{+2.53}_{-3.39}$ &$67.41^{+1.36}_{-1.13}$ &$-2.28^{+0.84}_{-0.92}$ &$-1.36^{+0.32}_{-0.30}$ &$-0.64^{+0.26}_{-0.27}$ &$-0.94^{+0.25}_{-0.25}$ &$-1.85^{+0.36}_{-0.35}$ &$0.43^{+0.04}_{-0.05}$ &$0.02196^{+0.00051}_{-0.00051}$ &-840.03 \\
		&$w$CDM &$68.42^{+1.21}_{-1.23}$ &$-$ &$-0.84^{+0.04}_{-0.04}$ &$-$ &$-$ &$-$ &$-$ &$0.30^{+0.01}_{-0.01}$ &$0.02196^{+0.00063}_{-0.00063}$ &-844.28 \\
		&$\Lambda$CDM &$72.24^{+0.66}_{-0.66}$ &$-$ &$-$  &$-$ &$-$   &$-$ &$-$  &$0.32^{+0.01}_{-0.01}$ &$0.02195^{+0.00060}_{-0.00059}$ &-848.55 \\
		\hline 
		&M &$74.05^{+2.80}_{-2.82}$ &$71.36^{+2.03}_{-2.01}$ &$0.33^{+1.54}_{-1.55}$ &$-1.78^{+0.41}_{-0.40}$ &$-0.31^{+0.36}_{-0.36}$ &$-1.68^{+0.35}_{-0.34}$ &$-0.59^{+0.78}_{-0.76}$ &$0.36^{+0.04}_{-0.04}$ &$0.02197^{+0.00051}_{-0.00050}$ &-24.68 \\
		&G &$76.29^{+3.38}_{-3.35}$ &$71.79^{+2.52}_{-2.42}$ &$-0.29^{+1.93}_{-2.00}$ &$-2.39^{+0.80}_{-0.78}$ &$0.24^{+0.64}_{-0.62}$ &$-2.06^{+0.52}_{-0.54}$ &$-0.18^{+1.31}_{-1.35}$ &$0.37^{+0.04}_{-0.04}$ &$0.02196^{+0.00052}_{-0.00053}$ &-24.91 \\
		{D+U} &H &$75.51^{+2.90}_{-3.51}$ &$68.38^{+1.75}_{-1.49}$ &$-2.07^{+0.84}_{-0.94}$ &$-1.60^{+0.30}_{-0.30}$ &$-0.56^{+0.25}_{-0.27}$ &$-1.40^{+0.24}_{-0.23}$ &$-1.08^{+0.40}_{-0.42}$ &$0.41^{+0.05}_{-0.06}$ &$0.02199^{+0.00050}_{-0.00051}$ &-23.75 \\
		&$w$CDM &$69.12^{+1.43}_{-1.47}$ &$-$ &$-0.86^{+0.05}_{-0.05}$ &$-$ &$-$ &$-$ &$-$ &$0.30^{+0.01}_{-0.01}$ &$0.02199^{+0.00061}_{-0.00063}$ &-26.81 \\
		&$\Lambda$CDM &$72.32^{+0.66}_{-0.67}$ &$-$ &$-$ &$-$ &$-$ &$-$ &$-$ &$0.31^{+0.01}_{-0.01}$ &$0.02196^{+0.00059}_{-0.00061}$ &-26.22 \\
		\hline
	\end{tabular}
	\label{Tab:app}
	\caption{ `D+P', `D+D', and `D+U' represents the data combinations of DESI + Pantheon, DESI + DESY5, and DESI + Union, respectively. `M', `G', and `U' represent the cases that the priors of $w(z)$ are derived using the $\rm Mat \acute{e} rn$-3/2 kernel, the Gaussian kernels, and the Horndeski theory. $\ln (E)$ denotes the Bayesian evidence corresponding to each case.}
\end{sidewaystable*}

\FloatBarrier
\clearpage

\end{appendix}
\end{document}